\documentclass[pss]{wiley2sp} 
\usepackage{dblfloatfix} 
\begin{document}

\title{Structure and energetics of embedded Si patterns in graphene}

\titlerunning{Embedded Si patterns in graphene}

\author{%
  Daryoush Nosraty Alamdary,
  Jani Kotakoski,
  Toma Susi\textsuperscript{\textsf{\bfseries \Ast}}}

\authorrunning{Nosraty Alamdary et al.}

\mail{e-mail
  \textsf{toma.susi@univie.ac.at}, Phone:
  +43-1-427772855}

\institute{%
  University of Vienna, Faculty of Physics, Boltzmanngasse 5, 1090 Vienna, Austria}

\received{XXXX, revised XXXX, accepted XXXX} 
\published{XXXX} 


\abstract{%
\abstcol{%
Recent experiments have revealed the possibility of precise electron beam manipulation of silicon impurities in graphene. Motivated by these findings and studies on metal surface quantum corrals, the question arises what kind of embedded Si structures are possible within the hexagonal lattice, and how these are limited by the distortion caused by the preference of Si for $sp^{3}$ hybridization. In this work, we study the geometry and stability of elementary Si patterns in graphene, including lines, hexagons, triangles, circles and squares. Due to the size of the required unit cells, to obtain the relaxed geometries we use an empirical bond-order potential as a starting point for density functional theory. Despite some interesting discrepancies, the classical geometries provide an effective route for the simulation of large structures.}}

\titlefigure[]{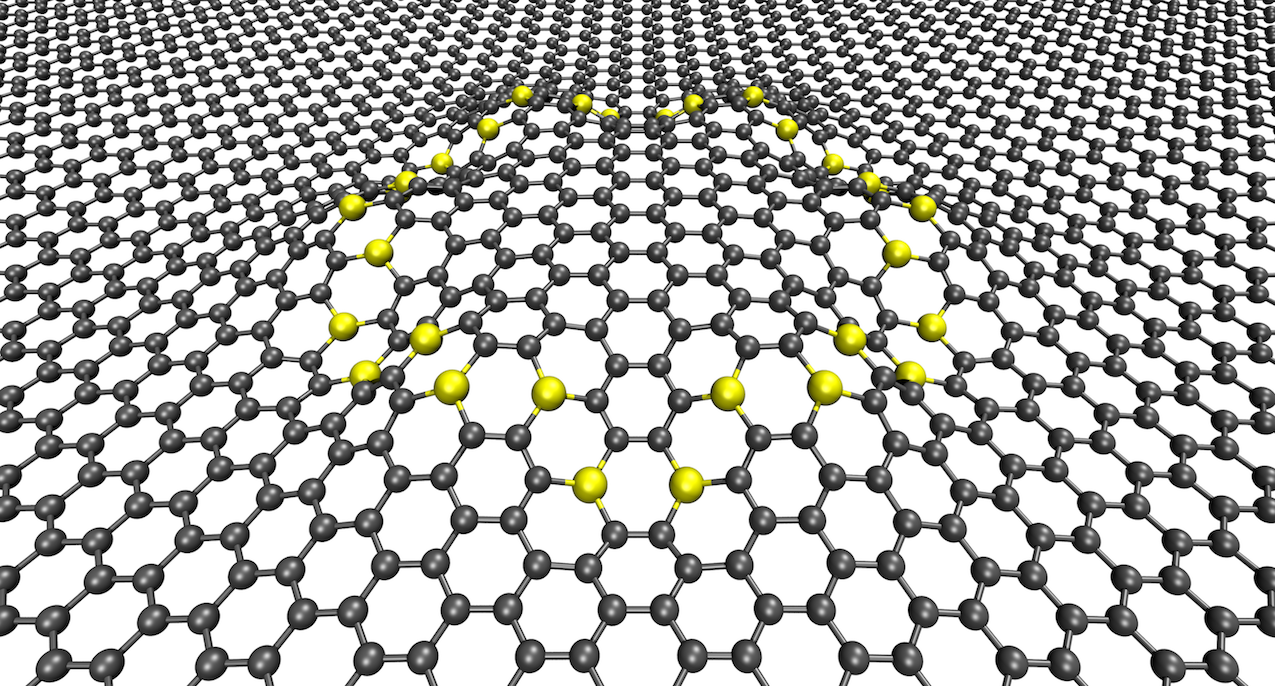}
\titlefigurecaption{A relaxed hexagonal structure of 30 Si embedded within the graphene lattice, containing in total of 1152 atoms in the unit cell. This pattern corrugates the graphene lattice in a symmetric way around a central plateau. 
}

\maketitle   

\section{Introduction}
Single-layer graphene not only has remarkable electronic~\cite{CastroNeto09RMP} and mechanical~\cite{Lee08S} properties, but it is also highly suitable for atomic-resolution transmission electron microscopy studies~\cite{Meyer08NL}. Due to its two-dimensional (2D) nature, each atom can be directly imaged, and the high conductivity reduces radiolysis and ionization, completely suppressing beam damage at electron acceleration voltages below 80 kV~\cite{Meyer12PRL,Susi16NC}. However, C atoms next to impurities such as Si heteroatoms embedded within the lattice~\cite{Ramasse13NL} are less strongly bound than atoms of the bulk~\cite{Susi12AN}. Scanning transmission electron microscopy (STEM) with 60 keV electrons cannot quite outright eject them, but instead induces out-of-plane dynamics~\cite{Susi14PRL} that allow the Si atoms to be non-destructively moved with atomic precision~\cite{Susi17UM}. These findings have raised the question of what kinds of stable patterns could be possible within the bounds of lattice symmetry.

\begin{figure*}[ht!]%
\includegraphics*[width=\textwidth,height=3.2 cm]{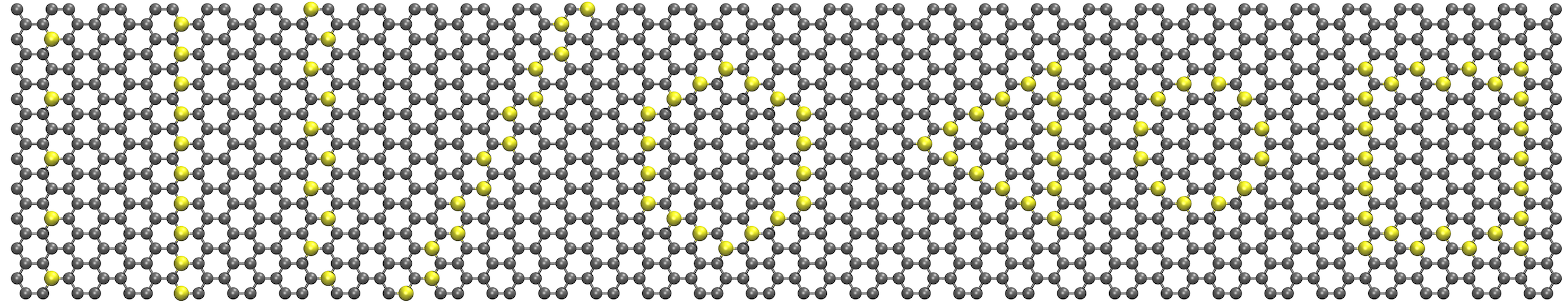}
\caption{An overview of the basic categories of structures studied in this work, from left: dashed zigzag (ZZ) line, ZZ line, A-B zigzag line (A-B ZZ), armchair (AC) line, hexagon, triangle, circle, square.}\label{Fig1}
\end{figure*}

Precisely designed Si structures could be of importance for at least two reasons. First, they raise the possibility of confinement of the graphene surface states, similarly to quantum corrals created by scanning tunnelling microscopy on metal surfaces~\cite{Crommie93S} since the early 1990s. Although embedded Si impurities certainly differ from adatoms on a metal surface, it is possible that closed rings or similar structures could also confine graphene electronic states into standing wave patterns. The second reason is the possible enhancement of graphene surface plasmons~\cite{Grigorenko12NP} near the impurities. Electron energy loss spectroscopy near single Si impurities has provided evidence for localized enhancement of the plasmon resonances~\cite{Zhou12NN}. Arranging many impurities into patterns whose dimensions match the plasmon wavelength might result in stronger antenna enhancement~\cite{Susi15FWF}, and their shape might allow plasmons to be directed in interesting ways~\cite{Spektor17S}.

Before such experiments can be realized, we need to know what kinds of Si patterns are possible. The hexagonal symmetry of graphene restricts the possibilities for placing Si atoms within the lattice, and the two symmetry directions, zigzag (ZZ) and armchair (AC), further limit the number of inequivalent patterns. Within these limitations, at least five categories of elementary structures appear possible, namely lines (both ZZ and AC), hexagons, triangles, circles, and squares, with the latter two being generally impossible to perfectly realize (Fig.~\ref{Fig1}).

Two further considerations are important. One is relative stability: C--C bonds are more stable than Si--C bonds, which in turn are more stable than Si--Si bonds~\cite{Shi15AN}. Thus while 2D silicon carbide is stable~\cite{Lin13JMCC} and indeed more stable than 2D silicene~\cite{Vogt12PRL}, it is significantly less stable and more reactive than graphene~\cite{Susi17arXiv}. Similarly for Si impurity patterns, Si--Si bonds increase the energy of the system (although it may still be stable~\cite{Susi14PRL}), as will bonds between Si and C. More importantly, though, since the beam manipulation method is based on the inversion of Si--C bonds~\cite{Susi14PRL}, neighbouring impurities are difficult to control. The second issue is computational: the size of the unit cell needs to be large enough so that the structures and the distortion they cause in the graphene lattice do not interact significantly with their periodic images. Even distortions caused by small vacancies in graphene become apparent only in simulations involving hundreds of atoms~\cite{kotakoski_atomic_2014}. For this reason, computationally efficient empirical bond-order potentials are required to relax structures with up to 1000 atoms.

Despite close similarities between the analytical potential and DFT, we do find some differences in the local geometries of the Si. These mostly subtle differences are not trivial, as they highlight the role of the electronic structure of materials in their local bonding and overall geometrical configuration. In few cases, this leads to surprisingly different overall shapes. In general, the analytical potential has a tendency to introduce stronger out-of-plane corrugation of the graphene sheet, whereas DFT consistently prefers flatter atomic arrangements.

\section{Methods}

The Atomic Simulation Environment (ASE)~\cite{Bahn02ASE} enables the efficient design and manual adjustment of atomic structures along with structure optimization (we used FIRE~\cite{Bitzek06PRL} for force minimization). To obtain potential energies and forces, this needs to be coupled to a calculator, either based on an analytical potential or density functional theory (DFT). For the classical calculations, we settled on the Erhart-Albe (EA)~\cite{Erhart05PRB} Si--C potential as implemented in the \textit{Atomistica} package~\cite{Pastewka13PRB}. For DFT, we used the grid-based projector-augmented wave code \textsc{Gpaw}~\cite{Mortensen05PRB,Enkovaara2010} with the PBE~\cite{PBE-GGA} functional and \textbf{k}-point spacings of less than 0.2~\AA$^{-1}$. For smaller systems we used a combination of plane wave (enabling a strain filter; cutoff energy 600~eV) and finite-difference (FD) modes (grid spacing 0.18~\AA), and for large ones, the highly efficient atom-based-orbital (LCAO) implementation~\cite{Larsen09PRB} with a polarized double-zeta basis.

For each structure type, we studied using the EA potential the influence of both the structure size (e.g. how large an area is delimited by the Si atoms) as well as the unit cell size (amount of graphene between the structures). After designing a Si structure into the lattice, we found the optimal unit cell size by scaling the structure separately in the $x$ and $y$ directions while relaxing the atomic positions, and selecting the minimum energy size. After this, the structure was further relaxed by a successively stricter three-stage iteration of a strain filter (minimizing the stress; for smaller structures using DFT) and force minimization (maximum forces $<$0.001~eV/\AA). We then took the converged size of each structure, scaled it by the difference between the DFT and EA equilibrium {C--C} bond length, relaxed each with LCAO-DFT, and finally converged the electron density and total energy using FD-DFT.

To estimate the smallest cell where the periodic images of the structures do not interact and to compare stability between structures, we calculate the embedding energy per Si atom as
\begin{equation}
\varepsilon = \frac{E_{tot}-N_{C}\mu_{C}}{N_{Si}}-\mu_{Si},
 \label{eq1}
 \end{equation}
where $\mu_{C}$ is the chemical potential of C (energy per atom in pristine graphene), $E_{tot}$ the total energy of the system, $N_{C}$ the number of the C atoms, and $\mu_{Si}$ the chemical potential of Si calculated for a single Si atom in vacuum (zero for any classical potential). The value of $\mu_{C}$ was calculated in EA and DFT respectively to be $-$7.374 and $-$9.223~eV, while the value of $\mu_{Si}$ in DFT is $-$0.805~eV. This energy becomes constant for sufficiently large unit cells.

\begin{figure*}[th!]
\includegraphics[width=17.1cm]{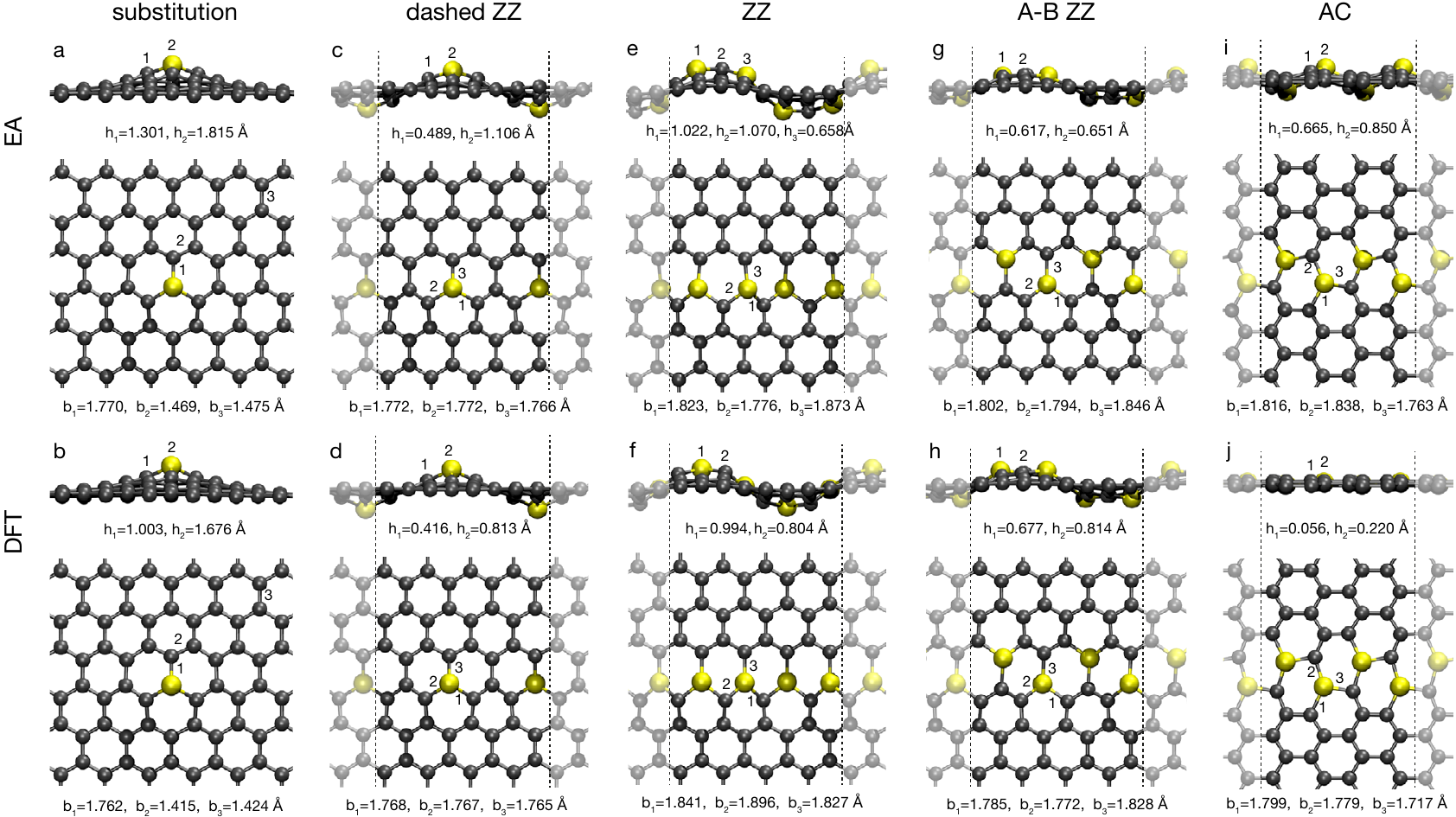}
\caption{Comparison of relaxed geometries with the Erhart-Albe potential (EA, top row) and density functional theory (DFT, bottom). a-b) Si substitution (crop of 160-atom unit cell), c-d) dashed zigzag line, e-f) zigzag line, g-h) A-B zigzag line and i-j) armchair line (each unit cell 48 atoms, lateral boundaries denoted by dashed lines). $\mathrm{h_i}$ denotes the height of atom $i$ from the lattice plane, and $\mathrm{b_i}$ the length of bond $\mathrm{i}$.}\label{comparison}
\end{figure*}

\section{Results}

We first compared the geometry a single trivalent Si substitution relaxed using the EA potential or with DFT (Fig.~\ref{comparison}a-b). The equilibrium graphene C--C bond length in EA is 3.6\% larger at 1.475~\AA\ (DFT: 1.424~\AA), while the Si--C bond is 1.770~\AA\ (DFT: 1.762~\AA). Thus graphene is slightly underbound with EA, whereas the Si atoms are comparatively overbound. The main difference, however, is the local corrugation: with EA, the second-nearest C neighbors buckle almost 0.3~\AA\ (8.3\%) further from the plane than with DFT, resulting in a Si height of 1.815~\AA\ (DFT: 1.676~\AA). Despite these discrepancies, the overall agreement is good. For multiple Si atoms in a unit cell, alternating their placement above and below the plane results in a lower energy~\cite{Susi14PRL}. For EA, the symmetric and antisymmetric energies are equal for cell sizes above 6$\times$ (number of pristine graphene hexagons between the Si), where the DFT energy is converged within 20 meV.

Next we turn to the different Si line structures (Fig.~\ref{comparison}c-j). Due to its simplicity (Si atoms on the same sublattice), the ZZ line is an elementary building block of most of the larger structures. In addition to a dense ZZ line, where every other C atom is replaced by Si, sparser arrangements such as the dashed ZZ line can be envisaged. Finally, an armchair line can be embedded into the lattice in the other symmetry direction, superficially resembling an A-B ZZ line where Si atoms are placed on alternating sublattices.

Most relaxed line structures are very similar between EA and DFT, with the exception of the AC line. Here EA predicts a significant out-of-plane corrugation (Fig.~\ref{comparison}i), while DFT finds a nearly flat structure (Fig.~\ref{comparison}j). For the ZZ line, EA yields a corrugated structure (Fig.~\ref{comparison}e), which is also reproduced by DFT (Fig.~\ref{comparison}f) when the EA configuration is used as the starting point for relaxation.

\begin{figure}[t]
\centering
\includegraphics[]{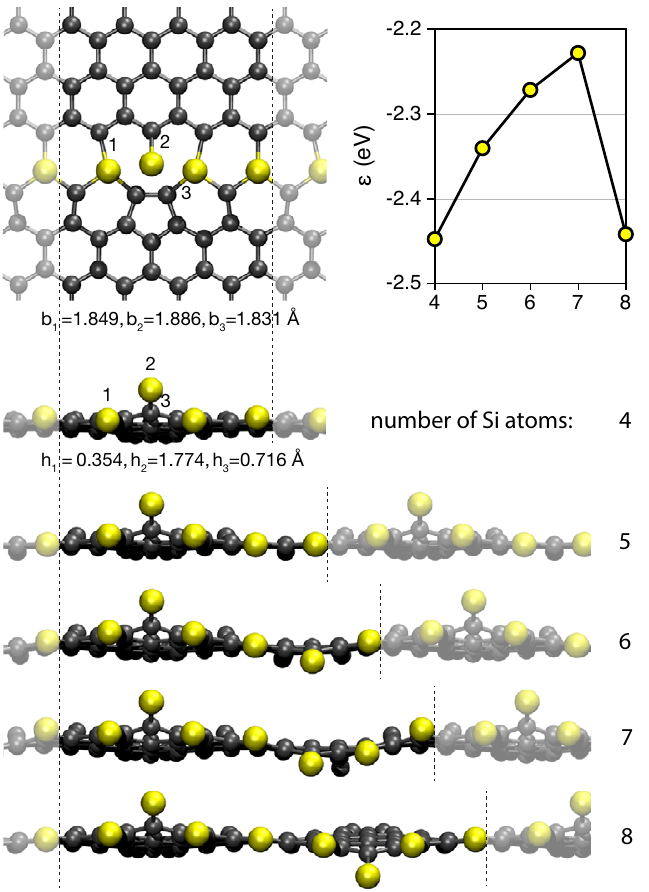}
\caption{The relaxed DFT structure and embedding energy of the flat ZZ line as a function of the unit cell size.}\label{ZZline}
\end{figure}

However, when initialized from a flat geometry, DFT can also yield another, rather surprising ZZ line: one of the four Si is lifted from the graphene plane, remaining bound to just one C atom with its other two C neighbours bonding in a pentagon (Fig.~\ref{ZZline}). The structure is otherwise almost completely flat---presumably to minimize the energy of the C atoms---reducing the embedding energy by 0.6~eV compared to the corrugated ZZ line. It appears that the DFT energy penalty of C atoms bound to two Si is so high that breaking two Si--C bonds to create just one more C--C bond becomes energetically favourable when a periodicity of at least four Si atoms is available in the cell. This can also be seen in the embedding energies plotted in Fig.~\ref{ZZline}, which show how the introduction of new in-plane Si atoms by increasing unit cell size increases the energy until the number of Si atoms reaches eight, which allows a second Si atom to buckle out of the plane. Before this, the separation of the singly-bound Si atoms is too short, which would result in too high strain between the C pentagons.

For the line structures, the embedding energy is converged to $<$10~meV when there are six rows of carbon between the Si lines. While a weak indirect interaction between the lines remains even at this distance, simulations with the system size doubled to introduce a second Si line into the unit cell show that the symmetry of the corrugation of the second line with respect to the first does not significantly influence the embedding energy. Since the second line neither leads to appreciable changes in the overall atomic structure, we limit the discussion here to the symmetric case possible in the smaller simulation cell.

\begin{figure*}[!t]
\centering
\includegraphics[width=10.5cm]{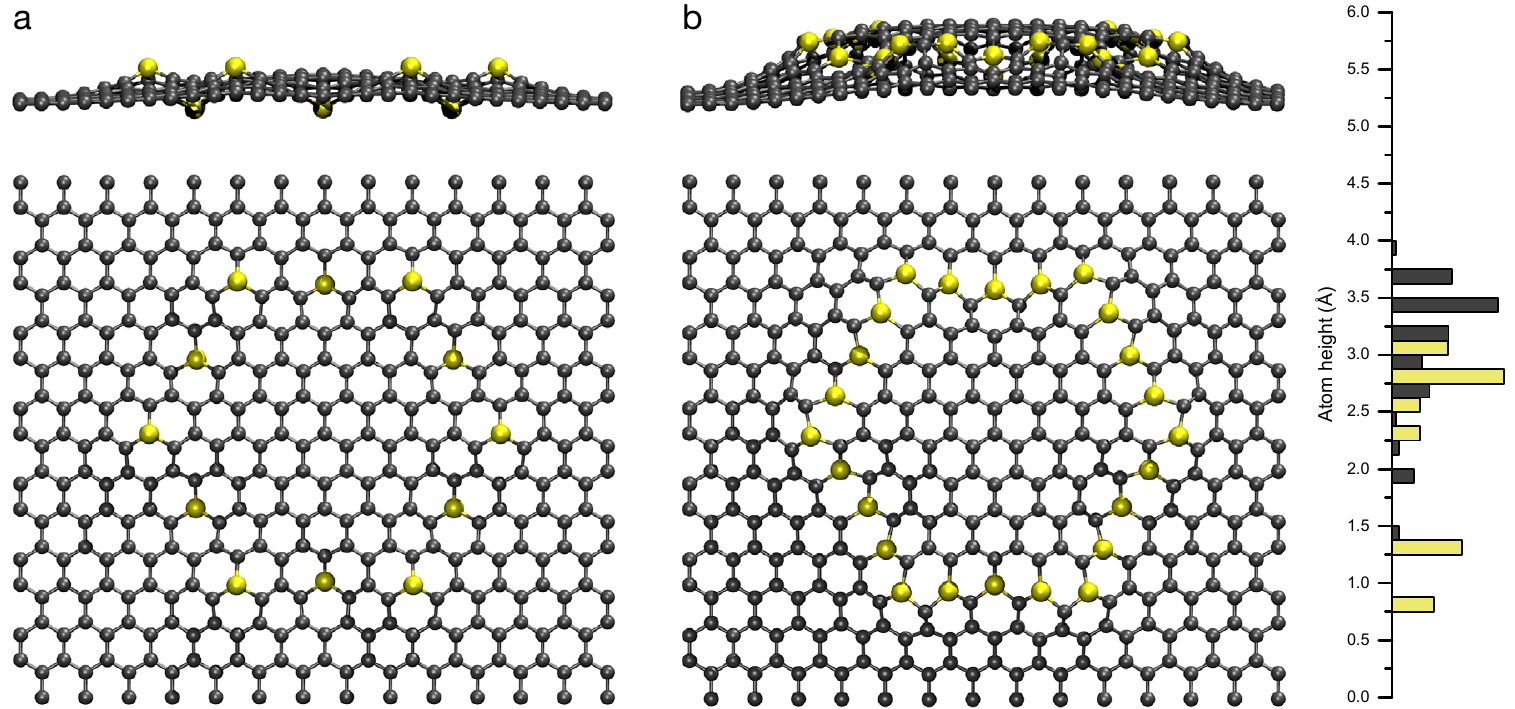}
\caption{The relaxed structures of the a) dashed and b) dense hexagon (including a histogram of atom heights). Dense Si lines result in a raised "mesa" delineated by the Si atoms that corrugate symmetrically around it at varying heights.}
\label{hexs}
\end{figure*}

\begin{figure*}[!b]
\centering
\includegraphics[width=17.1cm]{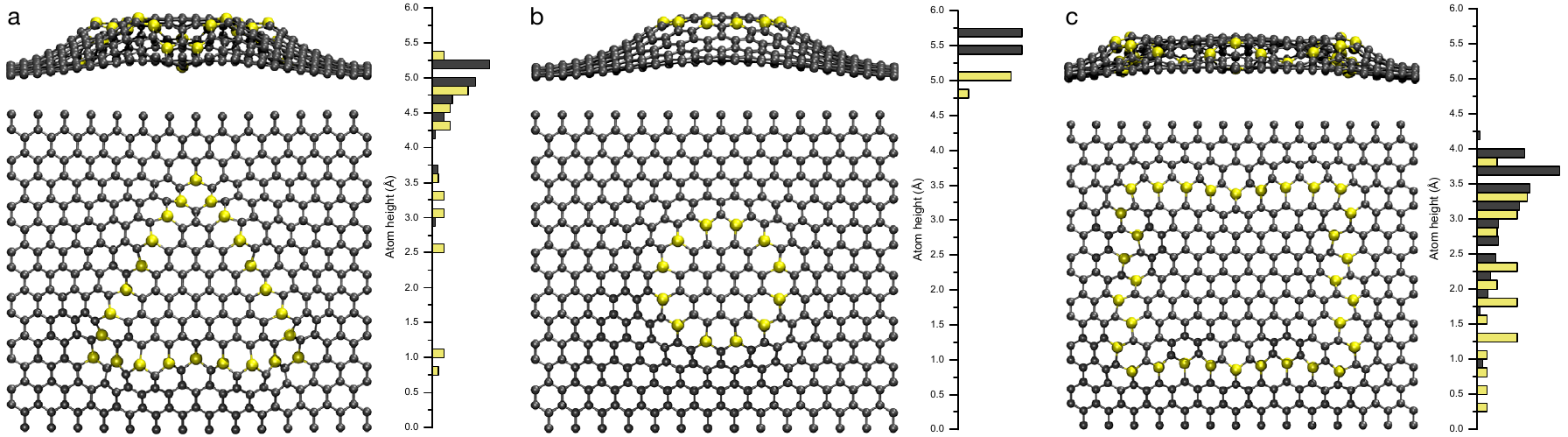}
\caption{The relaxed structures and histograms of atom heights of the a) triangle, b) small circle and c) square.}
\label{others}
\end{figure*}

In Table~\ref{energies} we have listed the embedding energy per Si as given by Eq.~\ref{eq1} for all of our structure classes using both the EA and DFT potentials. It is immediately obvious that bringing several Si atoms to close proximity reduces the embedding energy (thus stabilizing the structure), in some cases by more than 1~eV per Si. The embedding energies as calculated with the two methods show a trend that is similar to the atomic structures discussed above: the relaxed configurations that are alike between the two methods also show similar relative embedding energies. In fact, for most structures, the differences remain below 0.5~eV per silicon atom, the largest one ($\sim$1~eV) arising for the AC line with its EA out-of-plane corrugation.

\begin{table}[b]
\caption{The relative EA and DFT embedding energies per Si ($\Delta \varepsilon$; Eq.~\ref{eq1}) calculated for different structures with respect to the Si substitution ($\varepsilon^{sub}_\mathrm{EA} = -3.175$~eV, $\varepsilon^{sub}_\mathrm{DFT} = -0.780$~eV), and their difference ($\delta = \Delta\varepsilon_\mathrm{DFT}-\Delta\varepsilon_\mathrm{EA}$).}
\centering
\begin{tabular}[htbp]{@{}lccc@{}}
\hline
 Structure & EA (eV) & DFT (eV) & $\delta$ (eV) \\
\hline
 Dashed ZZ line 	& 0.022 & -0.197 & -0.219 \\
 ZZ line (corrug.) 	& -0.510 & -1.065 & -0.555 \\
 A-B ZZ line 		& -0.875 & -1.176 & -0.301 \\
 AC line 			& -0.274 & -1.335 & -1.061 \\
 ZZ line (flat) 	& --- & -1.665 & --- \\
\hline
 Dashed hexagon  	& 0.058 & 0.407 & 0.349 \\
 Hexagon  			& -0.492 & -1.098 & -0.606 \\
 Triangle  			& -0.527 & -0.856 & -0.329 \\
 Circle  			& -0.788 & -0.913 & -0.125 \\
 Square 		 	& -0.565 & -0.687 & -0.122 \\
\end{tabular}
\label{energies}
\end{table}

For the other pattern classes, we were mainly interested in large closed structures that could be reasonably simulated, and also potentially fabricated using electron beam manipulation. The first obvious closed structures are hexagons delineated by dense ZZ lines. Based on their EA energy convergence in terms of feature size and super-cell size, we settled on a square 7$\times$14 supercell of nominally 392 C atoms as a sufficient yet minimal graphene template. For the hexagon pattern, we considered both a dense 24 Si atom structure and a sparse version with only 12 Si, which we compare in Fig.~\ref{hexs}. The six ZZ lines are each $\sim$10 \AA\ in length, enclosing 27 full C hexagons within an area of $\sim$280 \AA$^2$, nearly comparable to graphene quantum dots that have been studied on metal surfaces~\cite{Hamalainen11PRL}.

The main difference between the two structures is the overall height of the "mesa" delineated by the Si atoms. The preferred local bonding of the Si within the line leads to their periodic up-and-down oscillation, similar to the corrugated lines. In the dense pattern, this seems only possible by raising the entire enclosed area several \AA\ above the lattice plane, whereas in the sparse hexagon, the Si atoms bond above and below the plane, greatly reducing the overall buckling. We also created sparser versions of the other structures, which exhibit the same kinds of differences and thus do not need to be discussed at length.

Like hexagons, triangles fully respect the lattice symmetry, with the 24-Si one shown in Fig.~\ref{others}a appearing qualitatively quite similar to the hexagon. Of structures that do not completely respect the symmetry, circles are particularly interesting. While it is not possible to make ones of arbitrary size that are fully circular, a 12-Si ring comes close and encloses exactly seven carbon hexagons (Fig.~\ref{others}b), \textit{i.e.}, embedded coronene. For this small structure where the Si atoms on the two halves of the circle occupy different sublattices, the "mesa" rising from the graphene plane is highly symmetrical and flat (consistent with smaller structures of other types). For larger circles, the Si atoms at the circumference weave up and down similar to the hexagon. Lastly we made structures as similar to a square shape as possible. While these also do not respect the hexagonal lattice symmetry, it is possible to make them using ZZ lines in one direction and AC lines in the other (Fig.~\ref{others}c), arguably making them an elementary shape.

\section{Discussion}
The easy availability of carefully parametrized and extensively tested bond-order potentials for C and Si makes it straightforward to simulate Si patterns in graphene (which is not the case for another interesting heteroatom, phosphorus~\cite{Susi172DM}). However, although we have shown that the classical structures mostly provide a good starting point for DFT, one has to be careful to avoid structurally and energetically distinct local minima. Starting from the corrugated EA-relaxed geometry of the AC line, DFT is able to find the correct flat structure. This is not the case for the flat ZZ line, which cannot be reached from the corrugated starting point.

In terms of the general differences in the relaxed structures, DFT tends prefer more symmetric arrangements with flatter graphene areas, perhaps due to the short range of the bond-order potential. However, the overall agreement of both the geometries and the relative energetics is surprisingly good. Unfortunately, the same does not extend to the simulation of electron irradiation: we found that EA gives qualitatively wrong results for the ejection of C neighbors to the Si, failing to reproduce the Si--C bond inversion~\cite{Susi14PRL} underlying the mechanism of electron beam manipulation.

Nonetheless, for obtaining relaxed geometries, starting with the EA potential offers a dramatic speedup. For example, the initial classical relaxation of the large hexagon takes only 10 min on single processor, whereas even with optimized LCAO calculator parameters, each of the subsequent relaxation steps consumes on average over 2000 CPU-min (well over 140 000 CPU-h in total).

All of the above structures have been designed using the simple trivalent Si substitution. A planar tetravalent bonding configuration~\cite{Ramasse13NL}, with the Si bonding to four C atoms in a graphene divancy, is also possible despite being slightly higher in energy~\cite{Susi14PRL}. However, while it would certainly reduce the corrugation of the lattice, that configuration is not possible to manipulate~\cite{Susi17UM} and thus is of little practical interest to us.

Finally, we should note that even larger structures can be designed and relaxed using the methodology of combining progressively more accurate simulations. For example, the unit cell with the 30-Si hexagon shown in the abstract figure contains 1152 atoms in total, yet it is even possible to obtain its high-quality wavefunctions due to the excellent parallelization of the \textsc{Gpaw} code. Unfortunately, despite modern computational resources, such system sizes are still prohibitively expensive for systematic studies, not the mention for simulating their electron beam stability and phonon modes using DFT~\cite{Susi16NC}.

\section{Conclusions}
We have presented here a multimodal approach for the efficient prediction of large embedded Si structures in graphene. After designing a Si pattern, its structure and unit cell size can be first roughly optimised using a classical bond-order potential. The structure can then be scaled to correct for the C--C bond length mismatch with respect to DFT, and then further relaxed using a computationally efficient atom-based orbital basis. Finally, the relaxed geometry can be used to obtain accurate energies, or indeed electron densities or wavefunctions, using an accurate real-space projector-augmented wave basis. This allows structures of many hundreds of atoms to be efficiently and accurately simulated, especially when all modes are implemented within the same simulation framework. The resulting large "quantum corral" structures will be interesting for their potential confinement of surface electronic states or plasmons, and thus make attractive targets for single-atom manipulation.

\begin{acknowledgement}
We acknowledge generous computational resources from the Vienna Scientific Cluster. T.S. acknowledges funding by the Austrian Science Fund (FWF) via project P 28322-N36 and J.K. by the Wiener Wissenschafts\mbox{-,} Forschungs- und Technologiefonds (WWTF) via project MA14-009.
\end{acknowledgement}

\providecommand{\WileyBibTextsc}{}
\let\textsc\WileyBibTextsc
\providecommand{\othercit}{}
\providecommand{\jr}[1]{#1}
\providecommand{\etal}{~et~al.}

\end{document}